\documentclass[a4paper,fleqn,usenatbib]{mnras}

\usepackage[T1]{fontenc}
\usepackage{ae,aecompl}

\usepackage{graphicx}	
\usepackage{amsmath}	
\usepackage{amssymb}	
\usepackage{lipsum}  

\newcommand{\bea}{\begin{eqnarray}}
\newcommand{\eea}{\end{eqnarray}}
\newcommand{\mrm}{\mathrm}

\title[Constraining modified gravity with BOSS]{The clustering of galaxies in the completed SDSS-III Baryon Oscillation Spectroscopic Survey: constraining modified gravity}

\author[E.-M. Mueller et al.]{Eva-Maria Mueller$^{1}$\thanks{E-mail: eva-maria.mueller@port.ac.uk}, Will Percival$^{1}$, Eric Linder$^{2,3}$, Shadab Alam$^{4,5,6}$, Gong-Bo Zhao$^{1,7}$, \newauthor{Ariel G. S\'anchez$^{8}$, Florian Beutler$^{1}$}
\\
$^{1}$ Institute of Cosmology $\&$ Gravitation, Dennis Sciama Building, University of Portsmouth, Portsmouth, PO1 3FX, UK \\
$^{2}$ Berkeley Center for Cosmological Physics $\&$ Berkeley Lab, University of California, Berkeley, CA 94720, USA \\
$^{3}$ Energetic Cosmos Laboratory, Nazarbayev University, Astana, Kazakhstan 010000 \\
$^{4}$ Department of Physics, Carnegie Mellon University, 5000 Forbes Avenue, Pittsburgh, PA 15213, USA \\
$^{5}$ McWilliams Center for Cosmology, Carnegie Mellon University, 5000 Forbes Ave., Pittsburgh, PA 15213 \\
$^{6}$ Institute for Astronomy, University of Edinburgh, Royal Observatory, Blackford Hill, Edinburgh, EH9 3HJ , UK \\
$^{7}$ National Astronomy Observatories, Chinese Academy of Science, Beijing, 100012, P. R. China \\
$^{8}$Max-Planck-Institut f\"ur extraterrestrische Physik, Postfach 1312, Giessenbachstr., 85741 Garching, Germany \\
}

\pubyear{2016}

\begin{document}
\label{firstpage}
\pagerange{\pageref{firstpage}--\pageref{lastpage}}
\maketitle

\begin{abstract}
We use baryon acoustic oscillation and redshift space distortion from the completed Baryon Oscillation Spectroscopic Survey, corresponding to data release 12 of the Sloan Digital Sky Survey, combined sample analysis in combination with cosmic microwave background, supernova and redshift space distortion measurements from additional spectroscopic surveys to test deviations from general relativity. We present constraints on several phenomenological models of modified gravity: First, we parametrise the growth of structure using the growth index $\gamma$, finding $\gamma=0.566\pm0.058$ (68\% C.L.). Second, we modify the relation of the two Newtonian potentials by introducing two additional parameters, $G_M$ and $G_L$. In this approach, $G_M$ refers to modifications of the growth of structure whereas $G_L$ to modification of the lensing potential. We consider a power law to model the redshift dependency of $G_M$ and $G_L$ as well as binning in redshift space, introducing four additional degrees of freedom, $G_M(z<0.5)$, $G_M(z>0.5)$, $G_L(z<0.5)$, $G_L(z>0.5)$. At 68\% C.L. we measure $G_M=0.980 \pm 0.096$ and $G_L=1.082 \pm 0.060$ for a linear model, $G_M= 1.01\pm 0.36$ and $G_L=1.31 \pm 0.19$ for a cubic model as well as $G_M(z<0.5)=1.26  \pm 0.32$, $G_M(z>0.5)=0.986\pm0.022$, $G_L(z<0.5)=1.067\pm0.058$ and $G_L(z>0.5)=1.037\pm0.029$. Thirdly, we investigate general scalar tensor theories of gravity, finding the model to be mostly unconstrained by current data. Assuming a one-parameter $f(R)$ model we can constrain $B_0< 7.7 \times 10^{-5}$ (95\% C.L).
For all models we considered we find good agreement with general relativity.
\end{abstract}






\section{Introduction}
For the last decade, increasingly accurate cosmological observations,
including the latest Planck datasets \citep{2016A&A...594A..13P} have reinforced a simple
cosmological model in which General Relativity (GR) describes all
gravitational interactions, about 70 per cent of the Universe's
current energy density is in form of a Cosmological Constant, and the
remaining 30 per cent is dominated by non-relativistic ``dark matter''
\citep[e.g.]{Weinberg:2012es}. While it is clear that the acceleration
mimics the cosmological constant in general effect, the exact physics
is unclear, and both new energy-density components and modifications
to GR, remain possibilities \citep{Copeland:2006wr,Koyama:2015vza}.

Observational effects of a dynamic energy-density component and
Modified Gravity (MG) are partially degenerate and careful data
analysis should take into account both possibilities.
 However, in general, observations of both
cosmological geometry and structure growth can distinguish between
these options as, in most modified gravity models, the growth of
structure is altered compared to general relativity. Purely
geometrical measurements, such as those from supernovae (SN) and baryon acoustic
oscillation (BAO) cannot
distinguish between these scenarios \citep{Huterer:2013xky}.

The breakdown of GR opens up a plethora of possible extensions, and no
unique physical direction for the modification has yet been
favoured. Consequently, recent analyses have focussed on generic
phenomenological descriptions, dependent on a small number of
parameters \citep{Daniel:2012kn, Asaba:2013mxj,Bean:2010zq,Zhao:2011te,Dossett:2011tn,Silvestri:2013ne}. These provide a mechanism to test for
particular types of behaviour, which, if detected, would provide
insight into the type of new physics required. Alternatively we can
think of these phenomenological models as providing complementary
tests of GR.

Galaxy redshift surveys provide a number of ways of obtaining
cosmological information by exploiting different physical mechanisms
that encode information in the observed distribution of galaxies. One
of the cleanest measurements is that of the Baryon Acoustic
Oscillation peak, observed within the clustering along (in
$\Delta z$) and across (in $\Delta \theta$) the line-of-sight. This
large-scale signal is difficult to distort by galaxy formation
processes, and allows robust measurements of the Hubble parameter $H$
and the angular diameter distance $D_A$ combined with the comoving sound
horizon $r_s$, which governs the primordial BAO position.

Galaxy surveys also allow measurements of the growth of structure via
Redshift-Space Distortions (RSD): the apparent clustering along the
line of sight receives a boost when redshifts are translated into
distances assuming all of the signal results from the Hubble
expansion, with amplitude proportional to the amplitude of
correlations in the peculiar velocity field. The amplitude of the
additive clustering signal is commonly parameterised by
$f(z)\sigma_8(z)$, where $f(z)$ is the
growth rate, and $\sigma_8(z)$ is the the linear-theory rms mass fluctuations in spheres of radius $8\,h^{-1}{\rm Mpc}$ 
 \citep{Song:2008qt}. Thus RSD provide a
measurement of the rate of growth of cosmological structure, which depends strongly on the large-scale strength of gravity.  A review of
BAO and RSD measurements is provided in \citet{2016arXiv160703155A}.

In this paper, we use the latest BAO and RSD measurements from the
Baryon Oscillation Spectroscopic Survey (BOSS; \citealt{Dawson:2012va}),
conducted as part of the Sloan Digital Sky Survey III (SDSS-III;
\citealt{Eisenstein:2011sa}), to test for evidence requiring modifications to
GR. The dataset used is described in \citet{2016arXiv160703155A}, and results
from the combination of a number of different measurements of BAO and
RSD determined using different methods. In particular, the
measurements are a combination of the BAO measurements of \cite{Ross:2016gvb} and \cite{2016MNRAS.tmp.1476B}
and the fits to the full clustering signal including RSD of \cite{2016arXiv160703150B}, \cite{2017MNRAS.464.1640S}, \cite{2016arXiv160703143G}
and \cite{2016arXiv160703148S}. These measurements are optimally combined using the method
described in \citet{2017MNRAS.464.1493S}, and are provided as correlated
measurements of $f\sigma_8$, $D_A/r_s$ and $Hr_s$ at three different
redshifts, $z=0.38$, $z=0.51$ and $z=0.61$, which we use along with
the $9\times9$ covariance matrix for these measurements. The RSD measurements are obtained under the assumption of a standard LCDM universe, which could potentially bias the results on more general theories of gravity. \cite{2016PhRvD..94h4022B} find, however, that within the context of MG models with scale-independent growth the constraints on $f\sigma_8$ are robust to these assumptions by applying the same analysis pipeline as was used in \cite{2017MNRAS.464.1640S} and \cite{2016arXiv160703143G} to mock catalogues of LCDM as well as the normal branch of DGP cosmologies. For models with a scale-dependent growth a pipeline which fully incorporated the MG model is preferable but beyond the scope of this paper. 

Our paper is presented as follows: In Section \ref{sec:Theory} we discuss the theoretical framework and common parametrisations of MG models. We focus on phenomenological descriptions of MG to connect fundamental theories to observations and to put general constraints on deviations from GR. A summary of the data sets used in this analysis can be found in Section \ref{sec:data_sets}. In Section \ref{sec:Analysis} we present the results of performing a Monte-Carlo Markov Chain (MCMC) analysis.



\section{Parametrising modifications to GR}
\label{sec:Theory}
In most theories of modified gravity the growth of structure is altered from GR, however, there is no unique description of the effect. Therefore we choose to parametrise deviation from GR in a phenomenological, model independent way. The following section summaries the parametrisations considered in this study.

\subsection{Growth index}
\label{sec:growth_index}
A minimal approach to model deviations from GR is to introduce one additional parameter to the $\Lambda$CDM model, parametrising the growth rate through the gravitational growth index $\gamma$ \citep{Linder:2005in,Linder:2007hg} as
\bea
f(a) = \Omega_m(a)^\gamma
\eea
with the scale factor $a$, $\Omega_m(a)=\rho_m(a)/[3M_p^2H^2(a)]$ where $\rho_m$ is the matter background density, $M_p$ the Planck mass, and $H(a)$ the Hubble expansion parameter. 
We also account for the contribution of $\gamma$ on RMS matter fluctuations today by rescaling $\sigma_8$ as
\bea
\sigma_{8,\gamma}(z) = \sigma_{8}(0)\frac{D_{\gamma}(z)}{D_{\mrm{GR}}(0)} \frac{D_{\mrm{GR}}(z_\mrm{hi})}{D_{\gamma}(z_\mrm{hi})} \eea 
with the growth factor calculated as
\bea
D_\gamma(a) = \mrm{exp} \left[-\int_a^1 da' f(a')/a' \right]
\eea
and assuming $z_\mrm{hi}=50$, well in the matter dominated era.

In GR we expect the growth index to be approximately constant with $\gamma \approx 0.55$. In this framework, the effect on the background expansion is treated separately from the growth of structure behaviour as an attempt to disentangle dark energy and modified gravity and to investigate the physical nature of extensions to the standard cosmological model. Its simplicity as well as its potential to differentiate between different models makes the growth index parametrisation an effective way to test deviations from GR against observations. However, potential scale dependent behaviour of modified gravity is not captured in this model.

Note, that the growth index can also be expressed in terms of modifications to the two Newtonian potentials. We will discuss this further in the next section.

\subsection{ $G_L$  and $G_M$ Parametrisation}
\label{sec:theory_GMGL}
In the Newtonian gauge perturbations to the metric can be described by the two gravitational potentials, $\phi$ and $\psi$,
\bea
ds^2 &=&a^2[-(1+2\psi)d\tau^2 + (1-2\phi)d\textbf{x}^2]
\eea
where $a$ is the scale factor, $\tau$ the conformal time and $\textbf{x}$ the spatial coordinate. Instead of phenomenologically modelling the growth of structure via the growth index, one can directly alter the evolution of the two gravitational potentials, $\phi$ and $\psi$, to account for potential modifications to GR.  
We can modify the Poisson equations
\bea
\nabla^2\psi &=& 4 \pi G a^2\rho\Delta \times G_{\mathrm{M}}  \label{eq:EGM}\\
\nabla^2(\psi+\phi) &=& 8 \pi Ga^2 \rho \Delta \times G_{\mathrm{L}} \label{eq:EGL}
\eea
introducing the dimensionless parameters $G_M$ and $G_L$. Here we have omitted the contribution of the anisotropic stress terms for simplicity since we are mainly interested in modifications to GR that arise in the matter dominated era. The standard GR perturbation equations are recovered for $G_M$=$G_L$=1. $G_M$ (short for $G_{\mrm{matter}}$) parametrises modifications to the growth of structure $\rho \Delta$ through the $\nabla^2\psi$ equation, whereas $G_L$ alters the lensing of light, $\nabla^2(\psi+\phi)$. This parametrisation of modified gravity has the advantage of allowing direct constraints on the fundamental, linearised perturbation equations as well as connecting to the cosmological observables while minimising degeneracies between the MG parameters. Note that, in the literature $G_M$ and $G_L$ is also referred to as $\mu$ and $\Sigma$, e.g. see \citet{Daniel:2012kn, Daniel:2010ky, Simpson:2012ra, Zhao:2011te,Song:2010fg}. 

Alternatively, one could also use the ratio of the two potentials, referred to as the gravitational slip,
\bea
\gamma_\mrm{slip}=\frac{\phi}{\psi}
 \eea
instead of $G_L$ to parametrise the modified Poisson equations. 

Measurements of $G_M$ and $G_L$ can be related back to specific MG theories as well as yielding implications for broad classes of theories. \cite{2016PhRvD..94j4014P} show that, for instance, Horndeski models seem to strongly favour deviations of $G_M$ and $G_L$ from unity to have the same sign.
In general, both MG parameters can be function of scale and redshift, $G_M(k,z)$ and $G_L(k,z)$. However, in this paper, we only consider redshift dependent behaviour, keeping both MG k-independent because of the current lack of a large set of scale dependent BOSS DR12 RSD measurements (but see \citep{Johnson:2014kaa,Johnson:2015aaa,Blake:2015vea} for other surveys). We model our ignorance of the exact redshift evolution of $G_M$ and $G_L$ in two ways: First, we assume a simple power law relation for both parameters
\bea
G_\mathrm{X} = 1 + (G_\mathrm{X}^{(s)}-1)a^s
\eea
with $X=\{M,L\}$, considering a constant redshift evolution (s=0), as well as a linear (s=1) and cubic (s=3) model. Here the subscript $(s)$ in $G_\mathrm{X}^{(s)}$ indicates the corresponding model.

While these parametrisations are not expected to reflect the actual evolution in many models, they can be viewed as providing a possible indication of deviations from general relativity. Note however that using a power law time dependence does not necessarily weight high and low redshift data correctly, and could bias the results \citep{Zhao:2011te}. Therefore we also consider other parametrisations below. If a signal is seen, then a wide variety of models or more detailed parametrisations should be employed.

Second, we bin $G_M$ and $G_L$ in two redshift bins, $z<0.5$ and $z>0.5$, adding four additional parameters to the standard $\Lambda$CDM model, 
\bea
\textbf{P}_{\mrm{MG}}=\{G_\mrm{X}(z<0.5),G_\mrm{X}(z>0.5)\}
\eea
with $X=\{M,L\}$.

We modify the publicly available MGCAMB code \citep{Hojjati:2011ix,Zhao:2008bn}, which itself is a modification of the Code for Anisotropies in the Microwave Background (CAMB) \citep{Lewis:1999bs}, to include these models. We assume deviation of GR arises during the matter dominated era, transitioning from the standard GR perturbation equation to the modified Einstein equation, as given by  Eq.~(\ref{eq:EGM}) and Eq.~(\ref{eq:EGL}), starting at redshift $z_\mrm{MG}<50$.

In the latter model, however, sharp transitions of $G_M$ and $G_L$ between the two redshift bins can cause numerical instability which leads to artificial constraints on $G_L$ from growth rate observations. We therefore smooth the transition between the bins using an arctan function of width $\Delta z = 0.002$. Note, that since MGCAMB evolves the perturbation equation using the $\mu$ - $\gamma_\mrm{slip}$ parametrisation we apply the smoothing to $\mu$ and $\gamma_\mrm{slip}$ with $\mu=G_M$ and $\gamma_\mrm{slip} = 2G_L/G_M -1$ in each bin respectively. 


The $G_M$-$G_L$ formalism can also be related to the growth index $\gamma$ (see Section \ref{sec:growth_index}). At sub-horizon scales $\gamma$ can be expressed in terms of $G_M$ following (32) of \citet{Pogosian:2010tj}
\bea
G_M = \frac{2}{3} \Omega_m^{\gamma-1} \left[\Omega_m^{\gamma} + 2 + \frac{H^\prime}{H} + \gamma \frac{\Omega_m^\prime}{\Omega_m} + \gamma^\prime \mrm{ln}(\Omega_m)\right]
\eea
 where primes indicate derivatives with respect to $\mrm{ln}a$. The $G_M$-$G_L$ formalism has the advantage of easily including observational constraints from CMB lensing, weak lensing or the ISW effect which are ignored when using the implementation outlined in Section \ref{sec:growth_index} which only accounts for direct growth rate measurements. Since the growth index only determines $G_M$, leaving $G_L$ or alternatively the gravitational slip $\phi_\mrm{slip}$ undefined, in order to fully apply this formalism, one needs to impose an additional theoretical prior on the model by fixing $G_L$ to unity (see e.g. \citealt{Simpson:2012ra}). Alternatively, one can fix the gravitational slip, $\gamma_\mrm{slip}=1$, as implemented in MGCAMB \citep{Hojjati:2011ix}. Beware, that these two approaches are essentially different parametrisations with different underlying theoretical assumptions and different observational effects. Therefore, we will refer to the growth rate parametrisation fixing $\gamma_\mrm{slip}$ as $\{ \gamma \ | \ \mrm{slip}\}$, and the parametrisation fixing $G_L$ as $\{ \gamma \ | \ G_{L} \}$.
 
 For more details on the relation between the different parametrisations in this framework see, for instance, \citet{Daniel:2010ky}.

\subsection{Scalar-tensor theories}
\label{sec:BZ}
Alternatively to a purely phenomenological description, one can start by considering first principals and a more general form of the Lagrangian to include a wide range of modified gravity models. Using symmetries, self-consistency conditions and stability requirements the Lagrangian can be simplified and a general expression for the perturbation equation can be derived.

Here we consider general scalar-tensor theories using the BZ parametrisation \citep{Bertschinger:2008zb, Zhao:2008bn},
\bea
G_M = \frac{1+\beta_1 \lambda_1^2 k^2 a^s}{1+\lambda_1^2k^2a^s} \\
\gamma_\mrm{slip} = \frac{1+\beta_2 \lambda_2^2 k^2 a^s}{1+\lambda_2^2k^2a^s} \label{eq:BZ}
\eea
with the gravitational slip, $\gamma_\mrm{slip}$, defined as 
\bea
 \gamma_\mrm{slip}=\frac{\phi}{\psi}
 \eea
and the dimensionless parameters $\beta_1$ and $\beta_2$ as well as the redshift evolution parameter $s$ and the length scale parameters $\lambda_1$ and $\lambda_2$.
 
 This parametrisation can capture the effect of most scalar-tensor theories in the quasi-static regime and can be used to test a wide range of modified gravity theories.
 
 A subset of this model can recover $f(R)$ theories: Assuming the relation 
 \bea
\beta_1^2= \frac{\lambda_2^2}{\lambda_1^2}=2-\beta_2^2  \frac{\lambda_2^2}{\lambda_1^2}
 \eea
 between the length scale parameters as well setting $\beta_1=4/3$ for a fixed coupling between the scalar field and matter, and $s\approx 4$ for viable models \citep{Zhao:2008bn, Hojjati:2012rf}, leaves us with a one parameter extension to GR given by
 \bea
 B_0 \equiv \frac{2H_0^2 \lambda_1^2}{c^2}.
 \eea



\section{Data sets}
In this section we outline the observational data sets used in our analysis, a combination of large scale structure (LSS) measurements, cosmic microwave background (CMB) experiments as well as supernovae type Ia (SN Ia) observations.
\label{sec:data_sets}
\subsection{BOSS DR12}
We use measurements of the post-reconstruction BAO position as a
function of direction to the line-of-sight, and the RSD amplitude,
measured from the Data Release 12 \citep{Alam:2015mbd} of BOSS. These
measurements were presented in \citep{2016arXiv160703155A}, and were obtained
by optimally combining measurements made using a number of methods
including measuring the BAO feature in the correlation function
\citep{Ross:2016gvb}, and power spectrum
\citep{2016MNRAS.tmp.1476B} multipoles, and RSD from fits to the shape of
multipole and angular wedge moments of the correlation function
\citep{2016arXiv160703148S,2017MNRAS.464.1640S}, and the power spectrum
\citep{2016arXiv160703150B,2016arXiv160703143G}. The methodology to derive the consensus constraints is discussed in detail in \citet{2017MNRAS.464.1493S}.

The galaxy catalogues used, and mitigation techniques for their
nuances are described in detail in \citep{Reid:2015gra}, which also
presents the targeting algorithm developed to select the galaxies: the
galaxies were selected from photometry taken using the Sloan
telescope \citep{Gunn:1998vh,Gunn:2006tw}, which was also used for
subsequent follow-up spectroscopy \citep{Smee:2012wd}. All the photometry
was re-processed and released in the Data Release 8 DR8
\citep{Aihara:2011sj}. Details of the spectroscopic data can be found in the DR12
data release paper \citep{Alam:2015mbd}, while the spectroscopic data
reduction pipeline and redshift determination are discussed in
\citet{Bolton:2012hz}.
\subsection{CMB}
We utilise the temperature $C_l^{TT}$, low-$l$ polarisation $C_l^{TE}$ as well as lensing $C_l^{\phi \phi}$ spectra from the Planck 2015 results \citep{2016A&A...594A..13P}. The constraints on modified gravity primarily come from lensing as well as the integrated Sachs-Wolfe (ISW) \citep{Sachs:1967er,Kofman:1985fp}. A more detailed summary about the effects of modified gravity on the CMB can be found in \cite{2016A&A...594A..14P}. 

\subsection{SN Ia}
\label{sec:data_sets_SN}
We use the joint light-curve analysis (JLA) of SN Ia observations by \citet{Betoule:2014frx}, a compilation of 740 SN Ia from the SDSS-II supernovae survey \citep{Frieman:2007mr,Kessler:2009ys, Sollerman:2009yu, Lampeitl:2009jq, Campbell:2012hi} as well as the Supernova Legacy Survey (SNLS, \citealt{Astier:2005qq, Sullivan:2011kv}) data.
Even though SN Ia observations cannot constrain the growth of structure directly, they provide strong constraints on the cosmological background parameters and hence decrease the overall uncertainty on all parameters.
\subsection{RSD measurements}
\label{sec:data_sets_RSD}
In addition to the DR12 BOSS data, we use RSD measurements from three different surveys (see Table \ref{tab:RSD}): the Six-Degree Field Galaxy Survey (6dFGS, \citealt{Beutler:2012px}), the SDSS Data Release 7 Main Galaxy Sample (MGS, \citealt{Howlett:2014opa}) and  the VIMOS Public Extragalactic Redshift Survey (VIPERS, \citealt{delaTorre:2013rpa}). 

The 6dFGS consists of 81,971 galaxies covering 17,000 $\mrm{deg}^2$ at low redshifts with $z_{\mrm{eff}}=0.067$. The growth rate measurement of $f\sigma_8=0.423\pm0.055$ of \citet{Beutler:2012px} was obtained modelling the 2D galaxy correlation function.
The MGS contains 63,163 galaxies distributed over 6,813 $\mrm{deg}^2$ at $z<0.2$ yielding to a growth rate of $f\sigma_8 = 0.49  ^{+0.15}_{-0.14}$, by fitting the two-point correlation function of galaxies in the sample.
VIPERS is a high redshift survey probing the LSS of the universe at $0.5<z<1.2$ covering 24 $\mrm{deg}^2$, measuring the growth rate $f\sigma_8=0.47\pm0.08$ using the monopole and quadrupole moments of the redshift-space correlations in their analysis. Table \ref{tab:RSD} summarises the RSD measurement used in this study. 

We do not use the BAO measurements of 6dFGS \citet{2011MNRAS.416.3017B} and MGS since these are likely correlated with the RSD measurement of the corresponding survey. Without a joint analysis of RSD and BAO measurements, or further assessment of the correlation, treating both measurements as independent could potentially lead to biased cosmological constraints. Therefore we only include the RSD measurements to get the best possible constraints on the growth of structure.

Similarly, due to the slight overlap with BOSS, we do not include the WiggleZ \citep{Blake:2011rj} and Data Release 7 LRG \citep{Samushia:2011cs} measurements, since both have much less precision than DR12.

\begin{table}
	\centering
	\caption{ Summary of the growth rate measurements used in this survey in addition to the DR12 BAO + RSD joint analysis.}
	\label{tab:RSD}
	\begin{tabular}{lc|cr} 
		\hline
		$f\sigma_8$ & $z_\mrm{eff}$&survey & reference \\
		\hline
		$0.423\pm0.055$ & 0.067 & 6dFGS & \citet{Beutler:2012px} \\
		\hline
		$0.49  ^{+0.15}_{-0.14}$ & 0.15 & MGS & \citet{Howlett:2014opa} \\
		\hline
		$0.47\pm0.08$ & 0.8 & VIPERS & \citet{delaTorre:2013rpa} \\
		\hline
	\end{tabular}
\end{table}



\section{Constraints on modified gravity}
\label{sec:Analysis}
In this section we perform a Monte-Carlo Markov Chain (MCMC) analysis using the publicly available CosmoMC \citep{Lewis:2002ah, Hojjati:2011ix} with our modifications to the code implemented as discussed in Section \ref{sec:Theory}. We run eight chains for each model until a convergence of $R-1<0.03$ is reached according to the Gelman-Rubin criterion \citep{Gelman:1992zz}. We assume a $\Lambda$CDM background with the MG parameters only affecting the perturbation equations. Therefore we vary the following cosmological parameters,
\bea
\textbf{P}=\{ w_{\mrm{cdm}},w_b, 100\theta_{\mrm{MC}},\tau, n_s, ln(10^{10}A_s),\textbf{X}_{\mrm{MG} } \}
\eea 
with the cold dark matter energy density $w_{\mrm{cdm}}=\Omega_{\mrm{cdm}}h^2$, baryon energy density $w_{b}=\Omega_bh^2$, the approximate sound horizon at last scattering $\theta_{\mrm{MC}}$ as used by CosmoMC, reionization optical depth $\tau$, scalar spectral index $n_s$, amplitude of the primordial curvature perturbations $A_s$ and the modified gravity parameters $X_{\mrm{MG}}$ for a given model. We fix the sum of the neutrino mass to 60 meV and assume an effective number of relativistic species $N_{\mathrm{eff}}=3.046$.
\subsection{Growth index}
\label{sec:Analysis_growth_index}
\begin{figure}
	\includegraphics[width=\columnwidth]{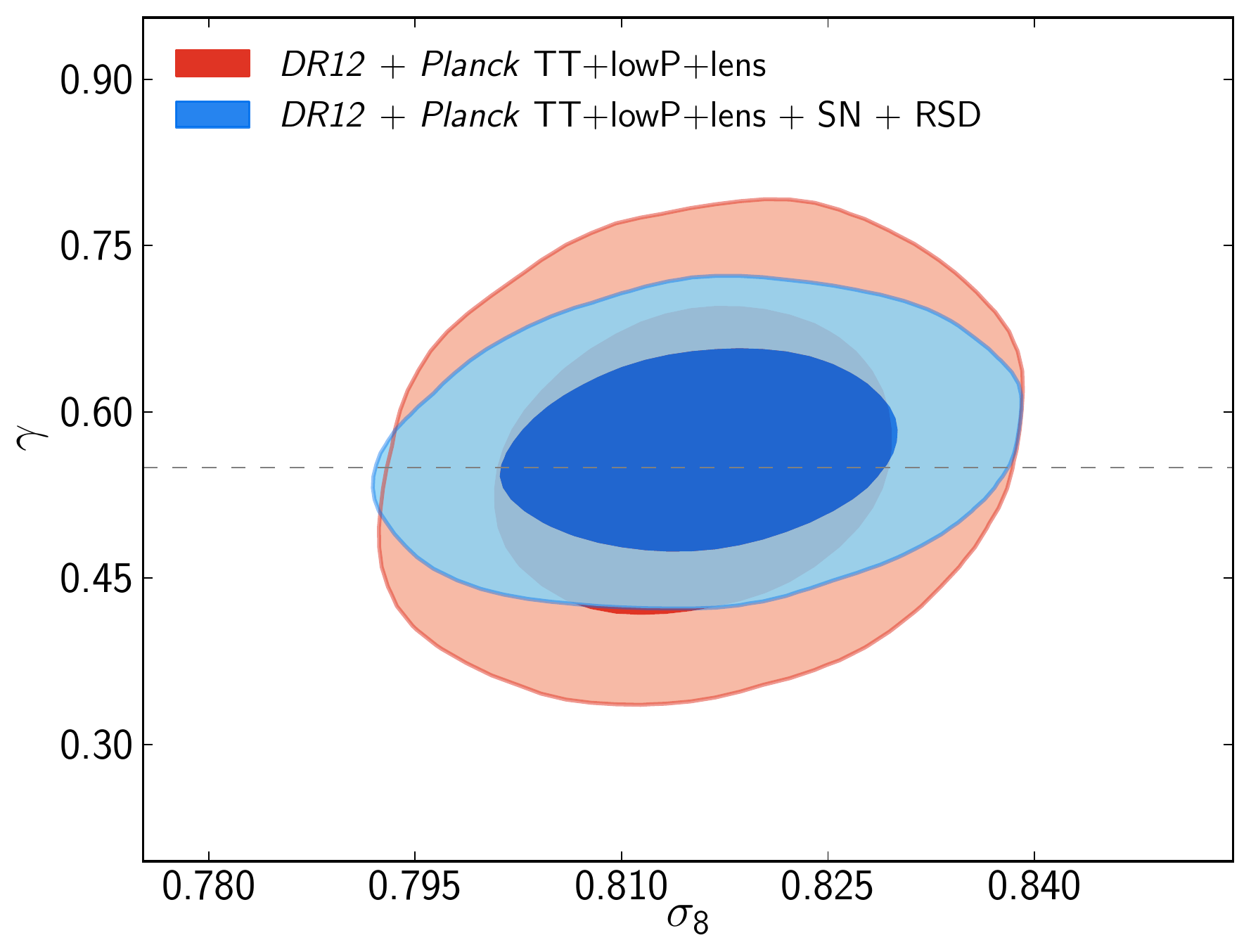}
  	\caption{68\% and 95\% constraints on the modified gravity parameter $\gamma$ and $\sigma_8$ in the base $\gamma\Lambda$CDM model, using the DR12 BAO+RSD combined analysis and Planck temperature,  low-$\ell$ polarisation and lensing (red contours), and including RSD measurement from additional LSS surveys as well as SN data as described in Section \ref{sec:data_sets} (blue contours). The dashed line shows the GR prediction for the growth index, $\gamma \approx 6/11$ \citep{Linder:2005in}.}
	\label{fig:s8_gamma}
\end{figure}

Fig.~\ref{fig:s8_gamma} shows the DR12 BAO + RSD consensus constraints in combination with other data sets (see Section \ref{sec:data_sets}) in the $\sigma_8-\gamma$ plane, adding the growth index $\gamma$ as a 1-parameter, modified gravity extension to the base $\Lambda$CDM. We find excellent agreement with GR measuring $\gamma=0.558\pm0.086$ at 68\% C.L. from the DR12 consensus measurements including Planck temperature compared to the 6-parameter base $\Lambda$CDM model. The improvement of the fit when varying $\gamma$ is marginal with $\Delta \chi^2 = 0.1$ compared to $\Lambda$CDM, albeit with higher complexity of the model. The constraints tighten when adding in further data sets yielding to a $10\%$ measurement uncertainty on the modified gravity parameter, $\gamma=0.566 \pm 0.058$, from DR12 BAO + RSD, CMB, SN and other RSD measurements. 


%


Our results are in good agreement with \citet{2017MNRAS.464.1640S}
who found $\gamma=0.609\pm0.079$ combining BOSS DR12 configuration space wedges measurements with Planck data as well as with 
\citet{2016arXiv160703143G}
 who quote $\gamma=0.52\pm0.10$ using Fourier wedges. $\gamma=0.52 \pm 0.10$ using Fourier space wedges. We can improve upon previous studies, i.e. \citet{Beutler:2013yhm} , by 30-40\%. 

As outlined in Section \ref{sec:Theory}, there is a subtlety in how the $\gamma$ formalism is applied when including effects of CMB lensing, weak lensing or the ISW effect. Instead of just parametrising the growth of structure (see Section \ref{sec:growth_index}), one can approximate the Newtonian potential $\psi$ in terms of $\gamma$ and evolve the modified perturbation equation as implemented MGCAMB, fixing either the ratio of the two potentials to unity, $\gamma_\mrm{slip}=1$ or fixing $G_L=1$; we denoted the former as the $\{\gamma \ | \ \mrm{slip}\}$ formalism and the latter $\{ \gamma \ | \ G_{L} \}$. We measure $\gamma=0.513 \pm  0.027 $ at 68 \% C.L in the $\{\gamma \ | \ \mrm{slip}\}$ parametrisation including effects of CMB lensing and the ISW effect.  Using the $\{ \gamma \ | \ G_{L} \}$ formalism we find $\gamma=0.529 \pm  0.067$ at 68 \% C.L . We find differences in the observational constraints because of the underlying theoretical assumption since the $\{ \gamma \ | \ G_{L} \}$ implementation leaves the gravitational lensing potential unchanged. Our measurement when including effects of CMB lensing, weak lensing or the ISW effect of the growth index is in good agreement with previous studies. For instance, \cite{Alam:2015rsa} found $\gamma = 0.477 \pm 0.096$ at 68\% C.L. for $\{\gamma \ | \ \mrm{slip}\}$  using CMASS DR11 and Planck 2013 angular power spectrum data and $\gamma=0.612\pm0.072$ using data from the Planck satellite in combination with six LSS surveys; \cite{Johnson:2015aaa} quote $\gamma=0.665\pm0.0669$ at 68\% C.L. using the $\{ \gamma \ | \ G_{L} \}$ parametrisation for a combination of multipole measurements from WiggleZ and BOSS, velocity power measurements from the 6dF survey as well as additional BAO, SN, CMB and ISW measurements.

 

\subsection{$G_M$-$G_L$ parametrisation}
\begin{figure}./
	\includegraphics[width=\linewidth]{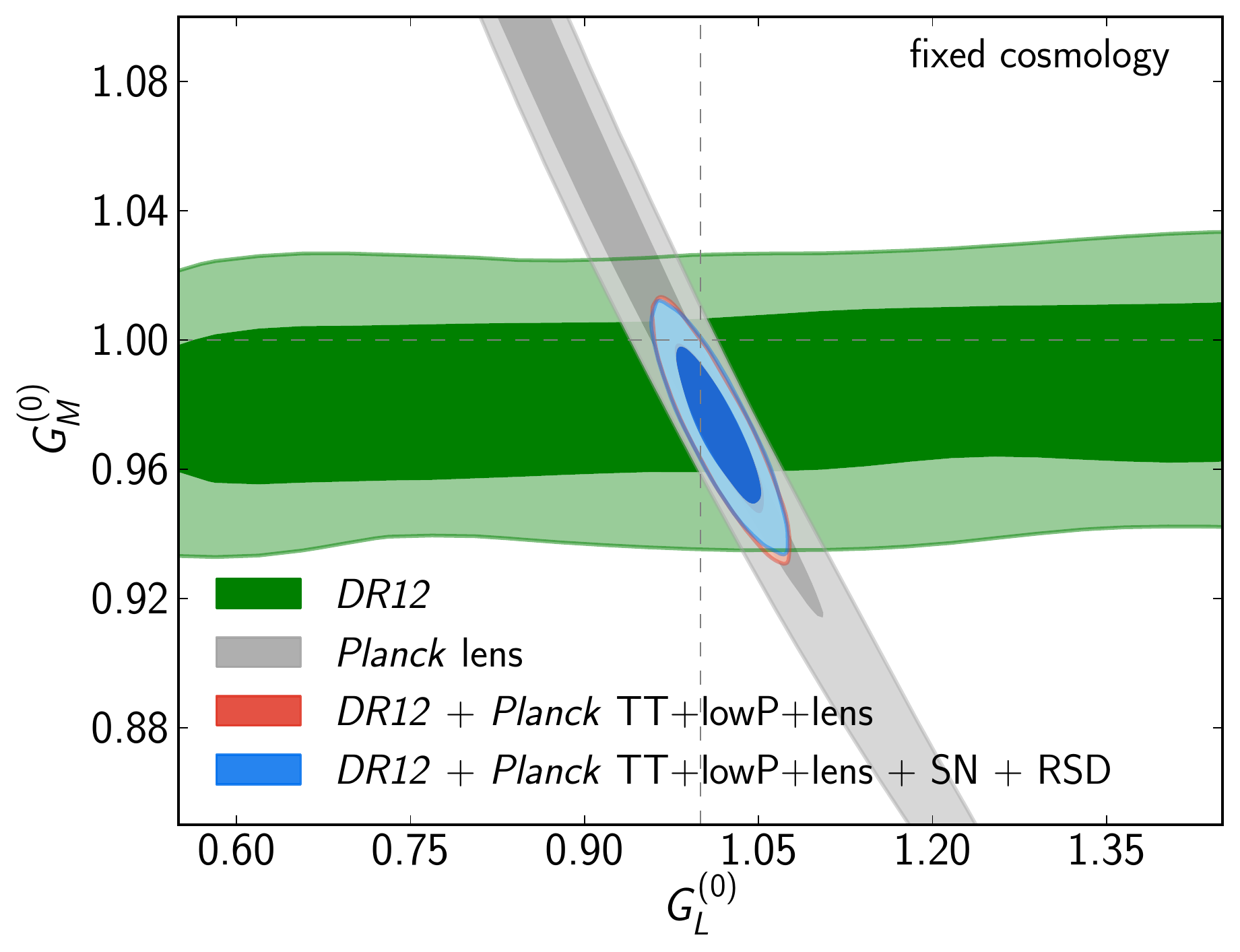}
  	\caption{Joint 68\% (dark shaded) and 95\% (light shaded) C.L. for the modified gravity parameters $G_{\mathrm{M}}^{(0)}$ and $G_{\mathrm{L}}^{(0)}$ assuming a constant model, as defined in Section \ref{sec:theory_GMGL}, for the different data sets: DR12 (green contours), Planck lensing (grey contours), DR12 + Planck temperature, low-$\ell$ polarisation and lensing (red contours) and DR12 + Planck temperature,low-$\ell$ polarisation and lensing + SN + RSD (blue contours) (for detail on the data sets see Section \ref{sec:data_sets}). Here we have fixed all other cosmological parameters to their Planck best fit value to highlight the degeneracies between the two modified gravity parameters; this figure should not be viewed as giving cosmological confidence regions. $G_M$ is mainly constrained by LSS RSD measurements whereas the uncertainty on $G_L$ is given by lensing measurements.}
	\label{fig:GMGLGs0_fixed}.
\end{figure}
\begin{figure*}
	\includegraphics[width=0.32\textwidth]{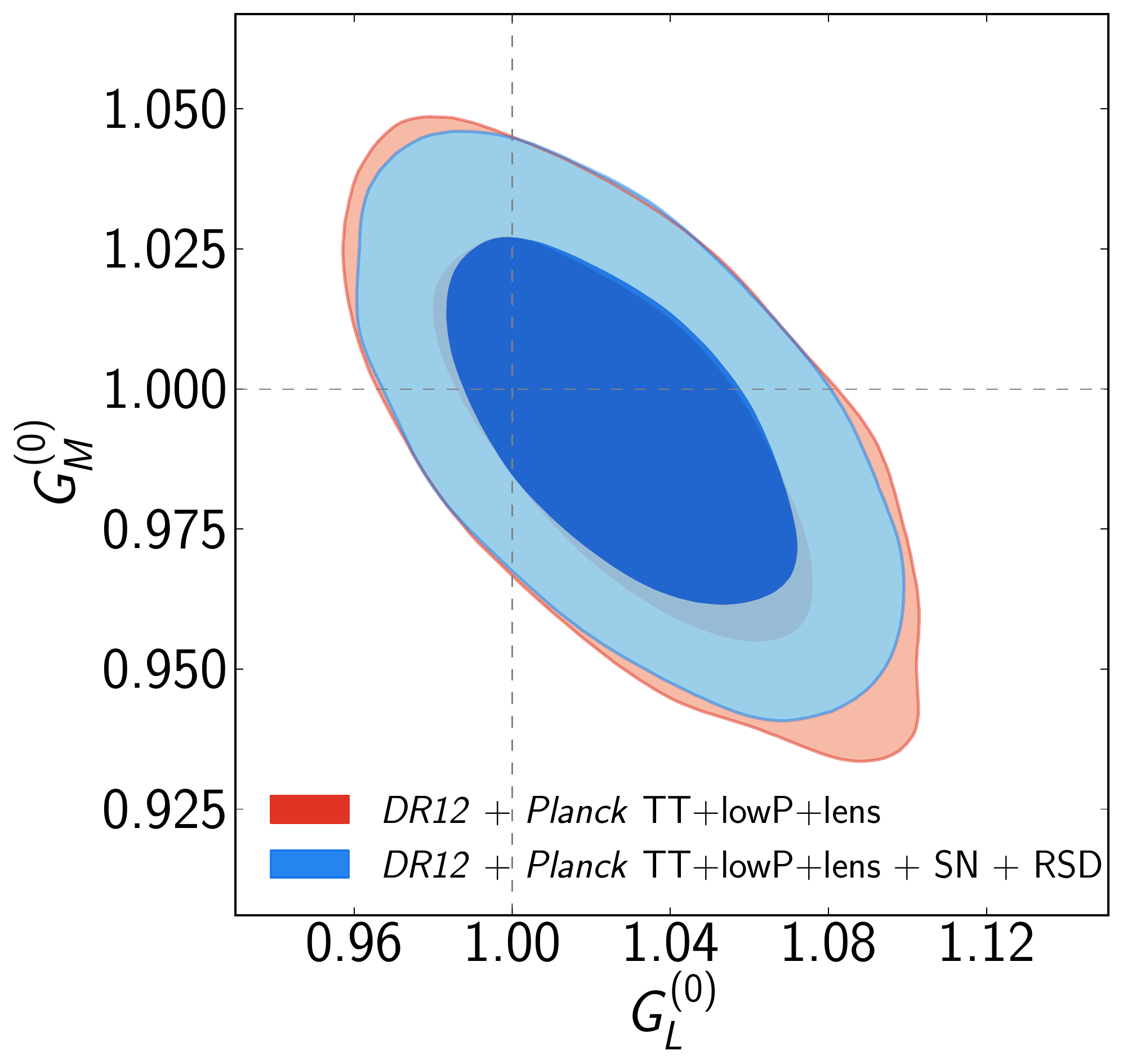}
	\includegraphics[width=0.32\textwidth]{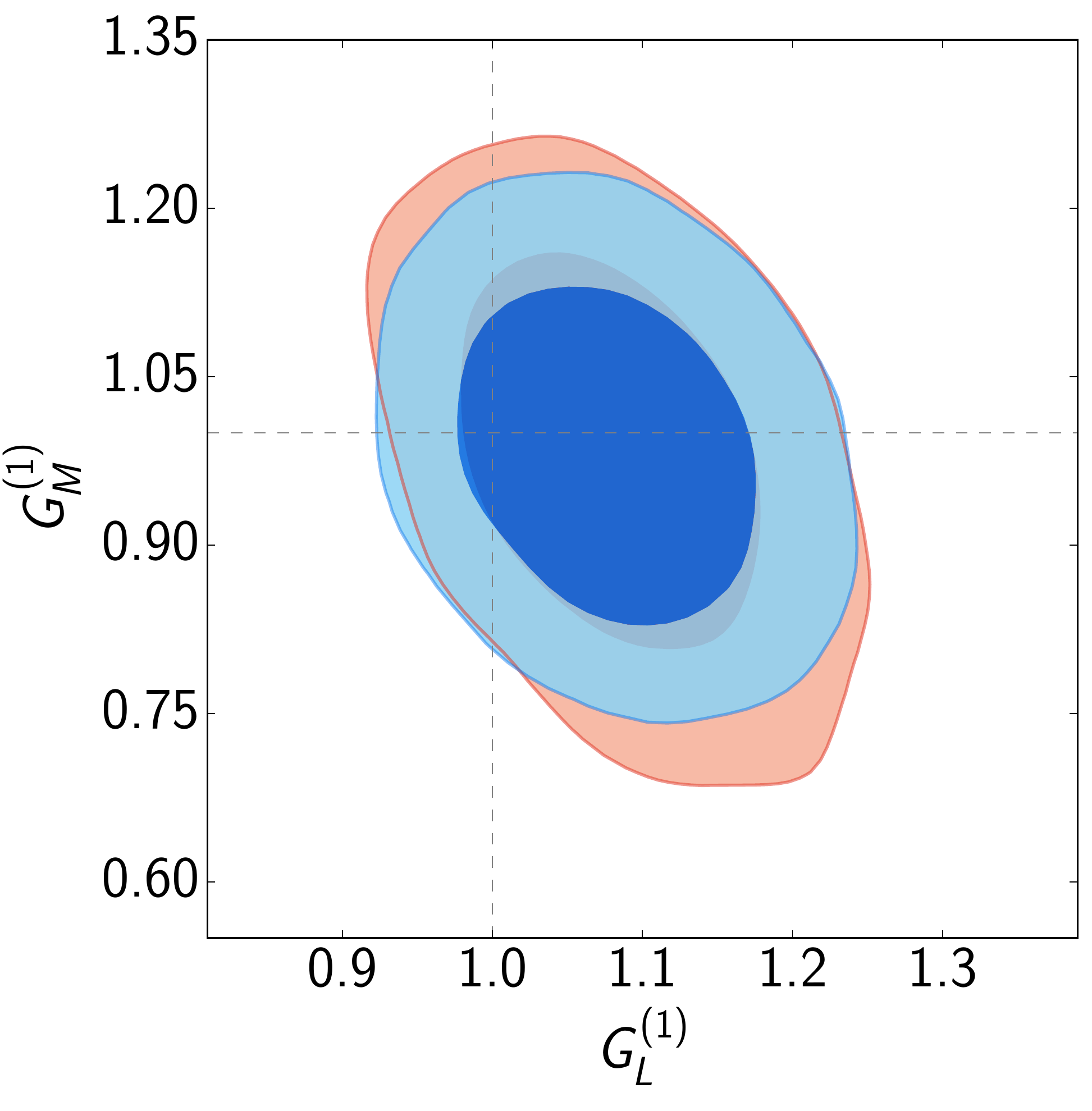}
	\includegraphics[width=0.32\textwidth]{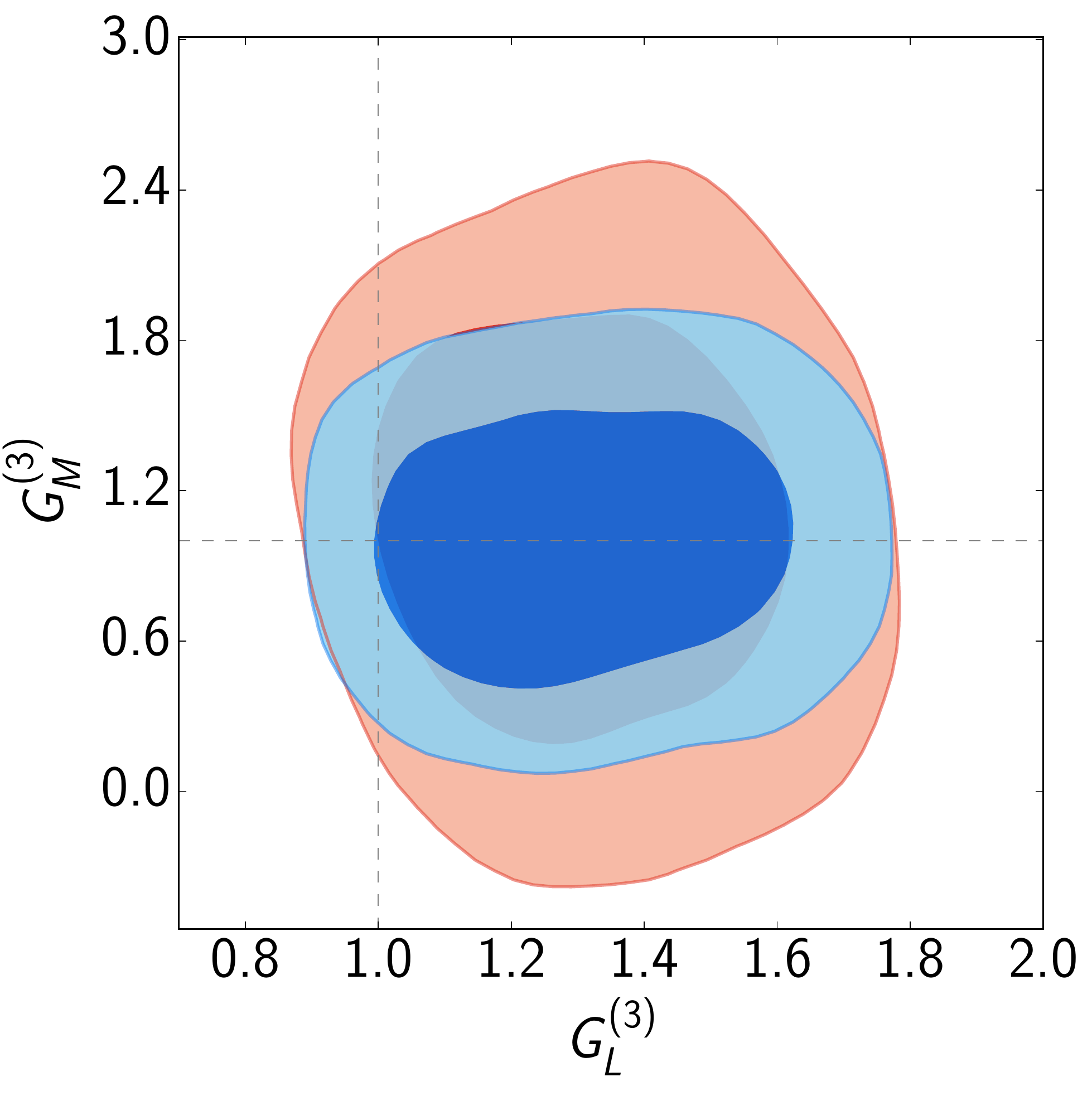}
  	\caption{68\% and 95\% confidence region of the modified gravity parameters $G_M$ and $G_L$ parametrised as $G_\mathrm{X}=1+(G_\mathrm{X}^{(s)}-1)a^s$, assuming a constant redshift evolution with $s=0$ [left panel], linear with $s=1$ [middle panel] and cubic with $s=3$ [right panel]. Note the very different scales. A stronger redshift dependency loosens up the constraints since the effect of modified gravity is diminished at high redshift, leaving the cubic model to be the least constrained scenario. In the constant model, however, deviations from GR start growing at high redshifts yielding tight constraints. The dashed grey lines show the GR prediction of the modified gravity parameters $G_M$=$G_L$=1.}
	\label{fig:GMGL_s}
\end{figure*}

Fig.~\ref{fig:GMGLGs0_fixed} shows the $68\%$ and $95\%$ confidence regions of $G_M$ and $G_L$ assuming a constant model (s=0) for different data sets. Here we have fixed the other cosmological parameters to highlight the degeneracy between the modified gravity parameters of the different cosmological probes. The DR12 combined sample BAO and RSD measurement can constrain $G_M$, whereas the uncertainty on $G_L$ is determined by CMB lensing in accordance to their definition (see Section~\ref{sec:theory_GMGL}). The combination of both growth of structure and lensing measurements, yields tight constraints on MG. 
The results of our MCMC analysis now marginalising over the $\Lambda$CDM cosmological parameters are displayed in Fig.~\ref{fig:GMGL_s}  and summarised in Table \ref{tab:GMGL}. Note that the 95\% C.L. tension with GR seen in Fig. 2 with fixed cosmology goes away when doing the proper marginalisation over cosmology. We find excellent agreement with GR for all models and data sets considered, finding both parameters to be unity within 1$\sigma$. The errors increase with a stronger redshift dependency since deviations from GR have a smaller impact at high redshift, i.e. $G_\mrm{X}^{(s)}$ contributes less to the overall $G_\mrm{X}$ defined as $G_\mathrm{X} = 1 + (G_\mathrm{X}^{(s)}-1)a^s$. Therefore, the constant model is the best constrained with the smallest uncertainty and the cubic model the least constrained with large errors. Note how assuming a particular redshift dependence can shift the contours. Including additional RSD measurements to the DR12 data set can improve the constraints, in particular for the cubic model, since additional measurements constrain the growth of structure over a larger redshift range. The $\chi^2$ for all three models is comparable, showing no preference for a particular redshift evolution, with $\Delta \chi^2 < 0.1$ compared to $\Lambda$CDM.

\begin{table}
	\centering
	\caption{Summary of the $68\%$ C.L. constraints on $G_M$ and $G_L$, marginalised over the $\Lambda$CDM parameters, from the MCMC analysis for a constant, linear and cubic model corresponding to the blue contours of Fig.~\ref{fig:GMGL_s}.}
	\label{tab:GMGL}
	\begin{tabular}{lcr} 
		\hline
		model & $G_M^{(s)}$ & $G_L^{(s)}$ \\
		\hline
		$s=0$: constant & 0.991 $\pm$ 0.022 & 1.030 $\pm$  0.030 \\
		$s=1$: linear & 0.980 $\pm$ 0.096 & 1.082 $\pm$ 0.060 \\
		$s=3$: cubic & 1.01 $\pm$  0.36 & 1.31 $\pm$ 0.19 \\
		\hline
	\end{tabular}
\end{table}

\begin{figure*}
	\includegraphics[width=\textwidth]{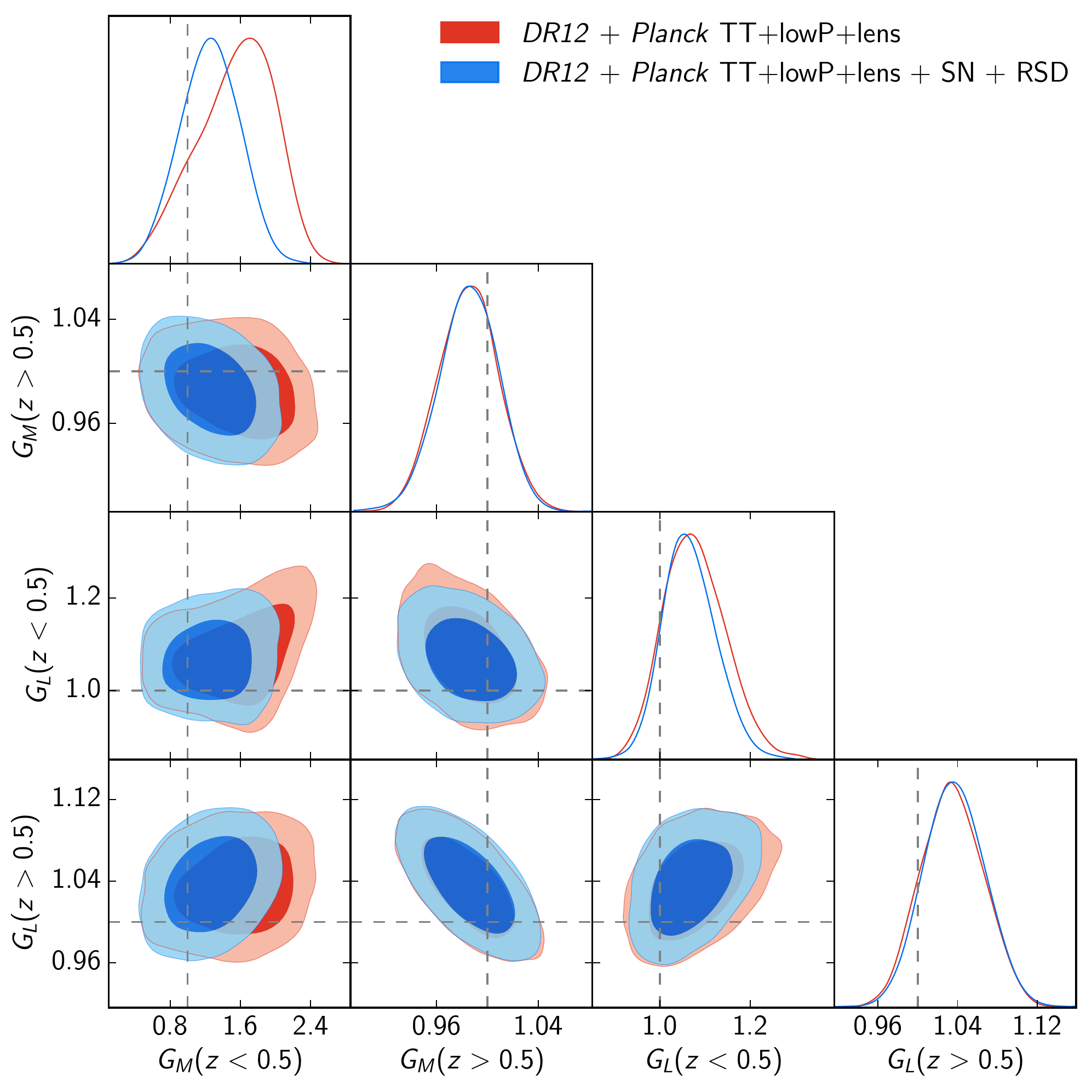}
  	\caption{68\% (dark shaded) and 95\% (light shaded) C.L. on the modified gravity parameters $G_M$-$G_L$ considering two redshift bins, $z<0.5$ and $z>0.5$ for different data sets: DR12 + Planck (red contours) and DR12+Planck+SN+RSD (blue contours). Including further LSS measurements to the DR12 sample can significantly improve the constraints on $G_M$ as the growth of structure is measured over a larger redshift range, especially to lower redshifts. However, the uncertainty on $G_L$ is dominated by CMB lensing and ISW measurements and therefore doesn't improve upon including additional RSD measurements. Dashed grey lines show the GR prediction, $G_M$=$G_L$=1.  }
	\label{fig:GMGL_bins}	
\end{figure*}

Secondly, we consider a model with the MG parameters binned in redshift space.
Fig.~\ref{fig:GMGL_bins} displays the constraints on $G_M$(z) and $G_L$(z) for two redshifts, $z<0.5$ and $z>0.5$, extending the standard $\Lambda$CDM model by a total of four extra parameters. The red contours show the uncertainty derived from the DR12 in combination with Planck temperature (TT), low-$\ell$ polarisation (lowP) and lensing data whereas the blue contours include additional SN and RSD measurements as described in Section \ref{sec:data_sets_SN} and Section \ref{sec:data_sets_SN}. The errors on $G_M$(z) are improved significantly by adding in additional growth rate measurements at multiple redshifts since $G_M$ alters the growth of structure. The improvements on $G_L$, however, are smaller because the constraints on $G_L$(z) are dominated by CMB lensing and the ISW effect. Using all datasets the 68\% CL results are
\bea
G_M(z<0.5)&=&1.26  \pm 0.32, \nonumber \\  
G_M(z>0.5)&=&0.986\pm0.022,  \nonumber \\
G_L(z<0.5)&=&1.067^{+0.050}_{-0.064}, \nonumber \\
G_L(z>0.5)&=&1.037\pm0.029 ,
\eea
 in very good agreement with GR at the 68\% CL. We find no significant improvement of the fit to the data compared to $\Lambda$CDM with $\Delta \chi^2 = 0.25$.
 
Our results are consistent with previous studies with slight differences arising due to the usage of different data sets: \cite{Johnson:2015aaa} derive constrains on $G_M$ and $G_L$, binned in both redshift and scale, using multiple measurements from the WiggleZ and BOSS DR11 CMASS and velocity power measurements from the 6dF survey in combination with CMB and SN data, confirming GR at 95\% C.L.. \cite{Song:2010fg} adopt a linear and cubic model for $G_M$ and $G_L$ with a combination of peculiar velocity and weak lensing measurements finding consistency with GR. For further studies see for instance \cite{Daniel:2010ky, Simpson:2012ra, 2016A&A...594A..14P}.

\subsection{Scalar-tensor theories}
\begin{figure*}
	\includegraphics[width=0.32\textwidth]{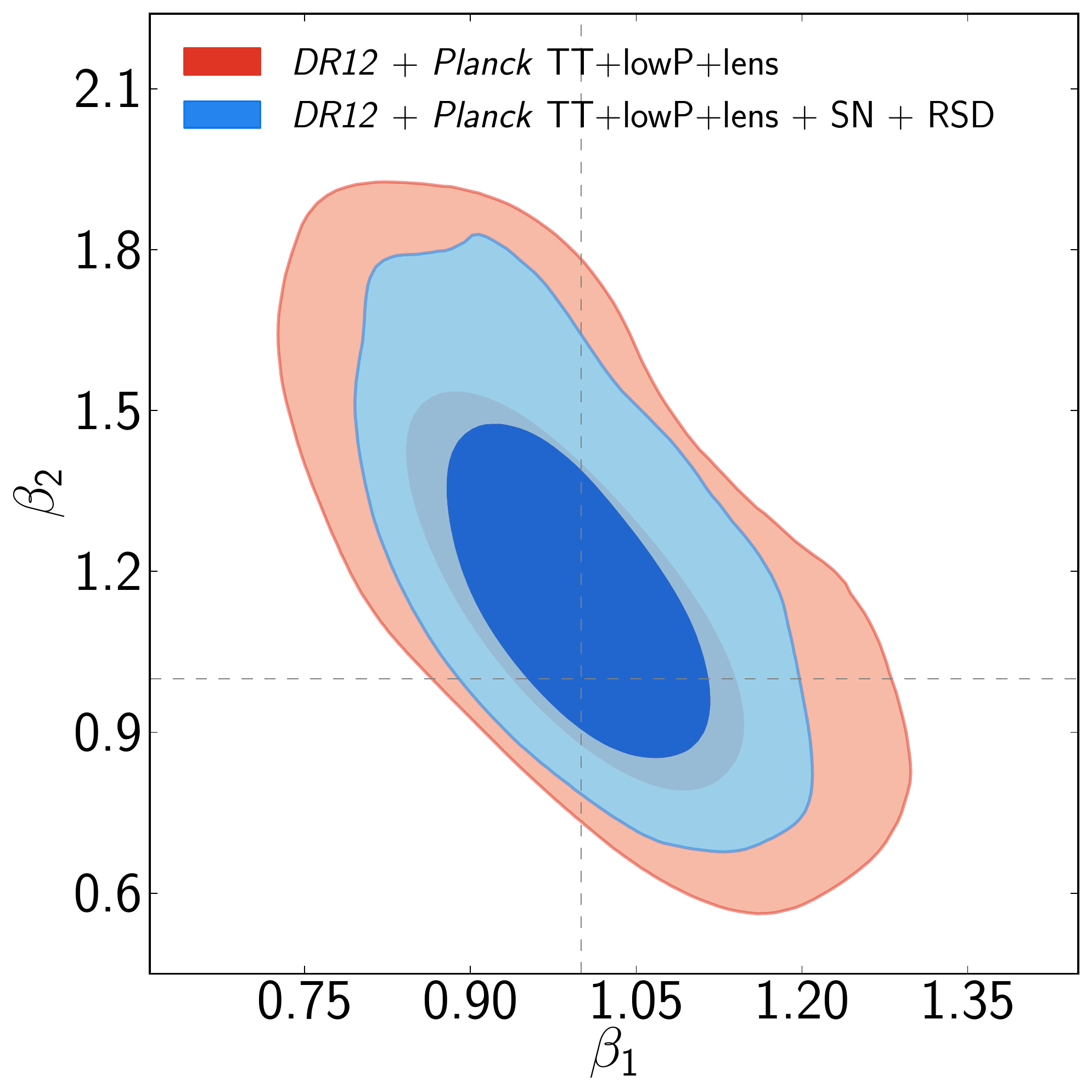}
	\includegraphics[width=0.32\textwidth]{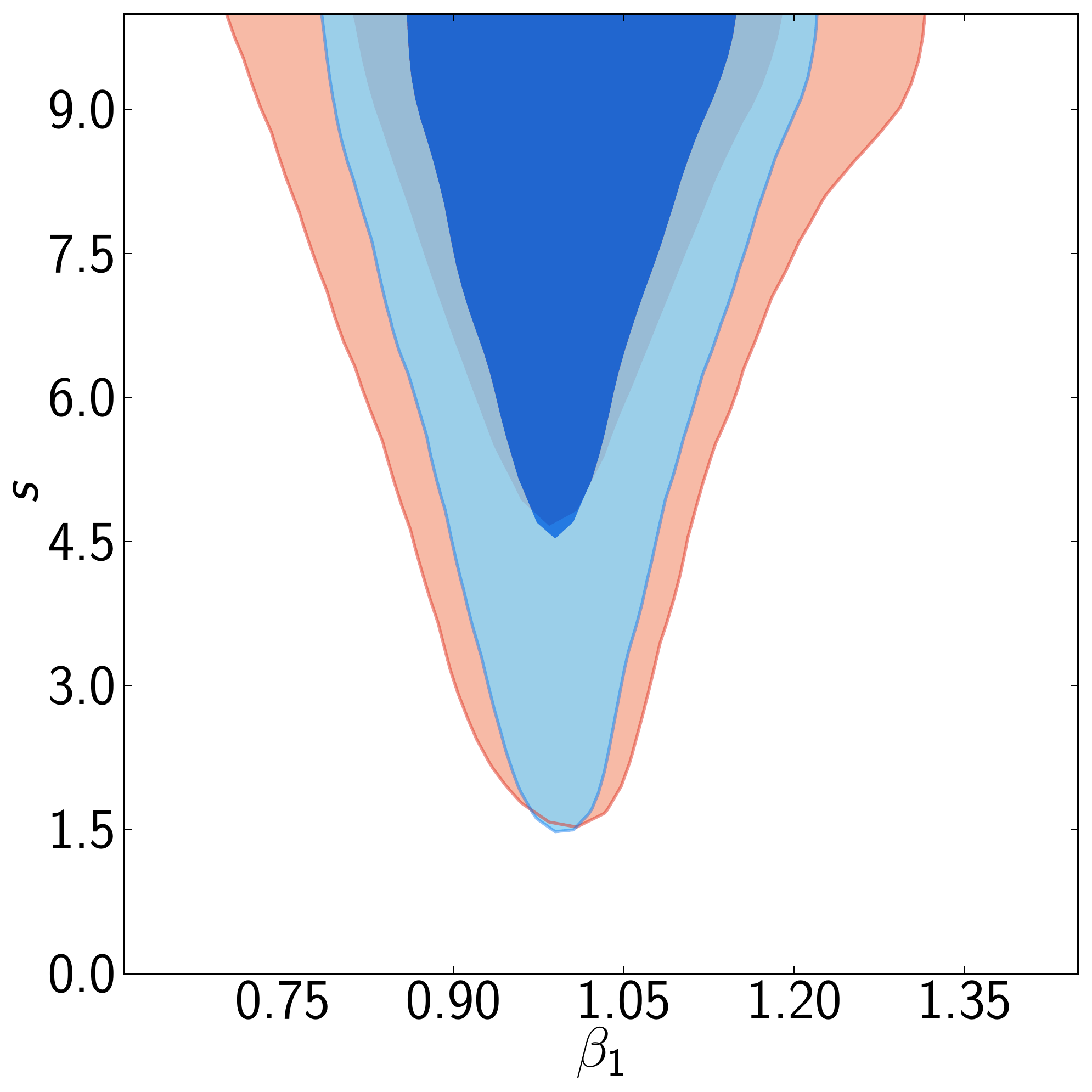}
	\includegraphics[width=0.32\textwidth]{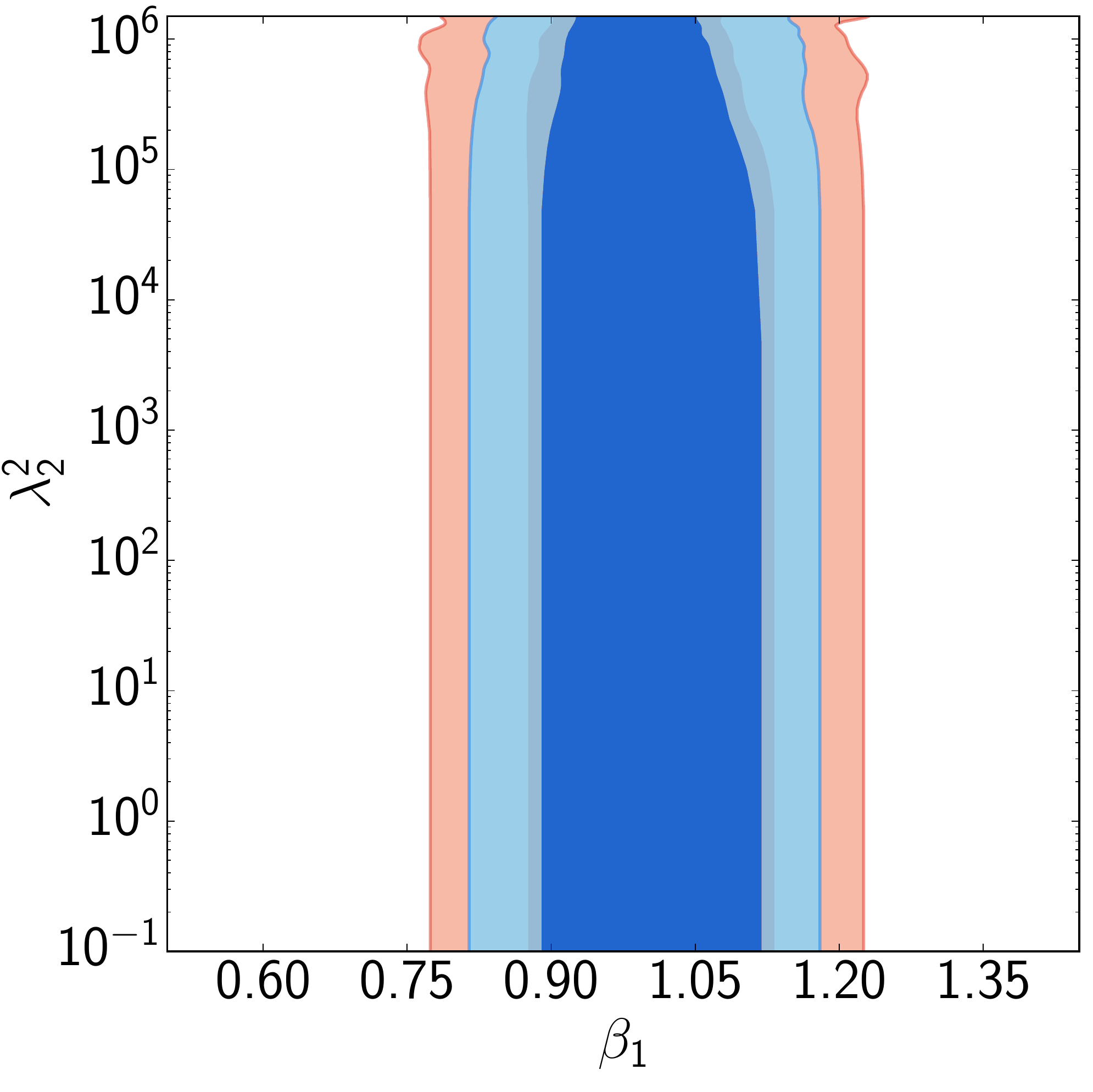}
  	\caption{2D contours of the BZ parameters, $\beta_1$-$\beta_2$ [left panel], $\beta_1$-$s$ [middle panel], $\beta_1$-$\lambda_2^2$ [right panel] for the DR12 combined analysis + Planck measurements (red contours) as well as including supernovae data and further RSD measurements as discussed in Section \ref{sec:data_sets} (blue contours). The constraints run into the upper limit of the prior on s; larger s would allow more extreme values of the other parameters, while smaller s would tighten the constraints (also see Fig.~\ref{fig:BZ_prior}).  }
	\label{fig:BZ}
\end{figure*}
\begin{figure}
	\includegraphics[width=\linewidth]{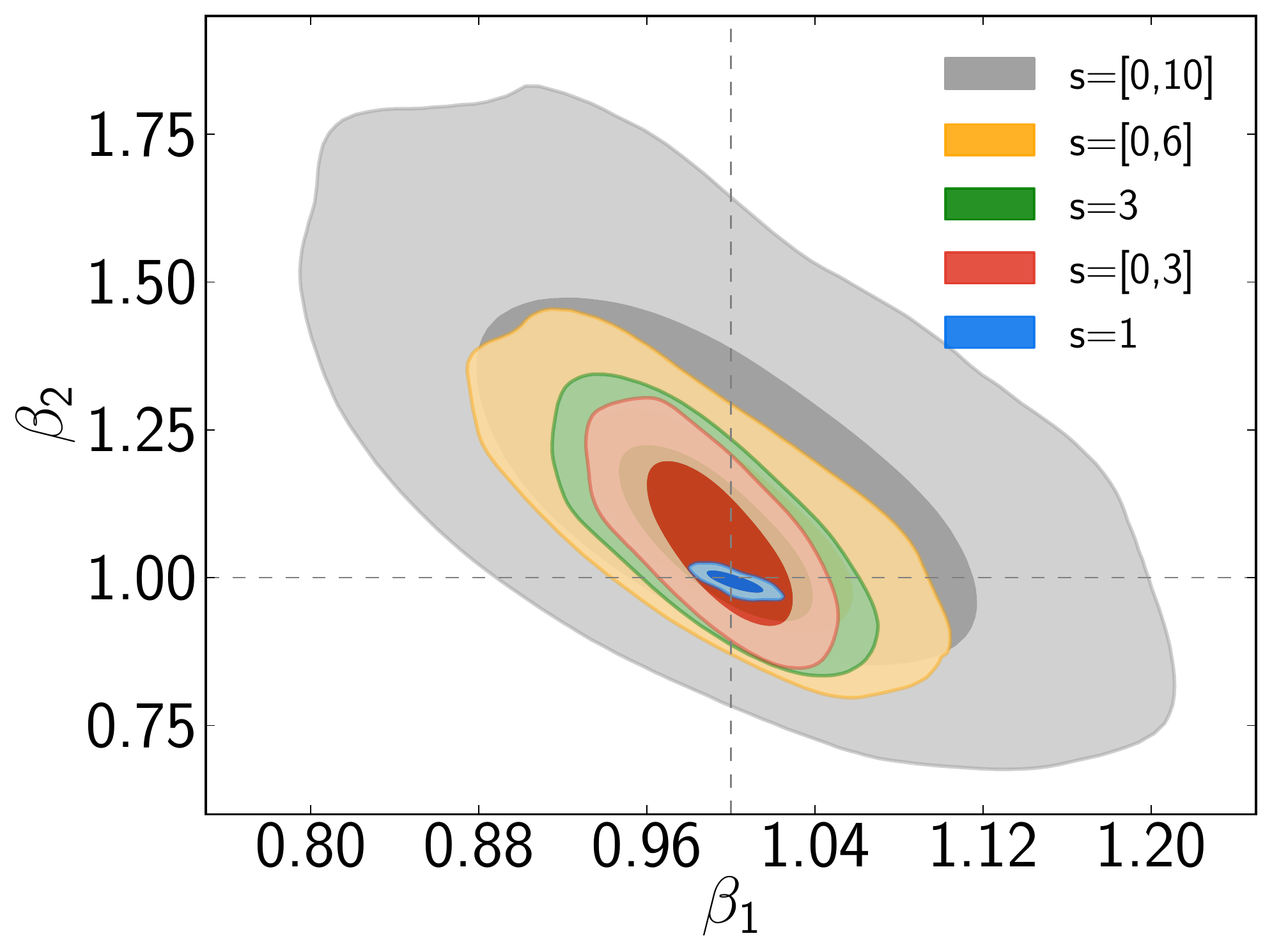}
 	\caption{2D contours of the BZ parameters, $\beta_1$-$\beta_2$, for different assumptions on $s$ using DR12+Planck+SN+RSD data;   grey contours assuming a prior range for $s$ of $s=[0,10]$, orange contours refer to $s=[0,6]$, red to $s=[0,3]$ while the green contours are for a model with $s$ fixed to 3, and blue for $s=1$.  }
	\label{fig:BZ_prior}
\end{figure}

\begin{table}
	\centering
	\caption{ Summary of the priors on the Scalar-Tensor theory parametrisation. All priors are linear.}
	\label{tab:priors}
	\begin{tabular}{lcc} 
		\hline
		Model & Parameter & Prior range  \\
		\hline
		BZ& $\beta_1$ & 0 - 3  \\
		&$\beta_2$ & 0 - 3  \\
		&$\lambda_1^2$ & (0 - $2)\times 10^6 \ \mrm{Mpc}^2$\\
		&$\lambda_2^2$ & (0 - $2)\times 10^6 \  \mrm{Mpc}^2$ \\
		&$s$ & 0 - 10  \\
		$f(R)$& $B_0$ & 0 - 0.01  \\
		\hline
	\end{tabular}
\end{table}
Fig.~\ref{fig:BZ} shows the likelihood constraints on the parameters of the BZ model, including a prior on $s$ given in Tab.~\ref{tab:priors}. As $s$ tends to infinity, we see from Eq.~(\ref{eq:BZ}) that the terms that depend on $\beta_1$ and $\beta_2$ become negligible, except at very low redshifts. Consequently in this limit, $\beta_1$ and $\beta_2$ can take any value without changing the model. If the data allows this limit, then we see that the $\beta_1$ and $\beta_2$ constraints are degenerate with the upper limit placed on $s$ by the prior. In effect we would only find meaningful constraints on $\beta_1$ and $\beta_2$ if we also measure $s$. Fig.~\ref{fig:BZ} shows that this is not the case for the data sets under consideration, and consequently, the constraints on $\beta_1$ and $\beta_2$ shown in the left panel of Fig.~\ref{fig:BZ}  are purely determined by the upper limit of the prior on s. Consequently, we do not quote any parameter measurements for this model. The constraints on $\beta_1$ and $\beta_2$ for different priors on $s$ can be found in Fig.~\ref{fig:BZ_prior}. Decreasing the prior range on $s$ reduces the uncertainty on $\beta_1$ and $\beta_2$ significantly.

\begin{figure}
	\includegraphics[width=\linewidth]{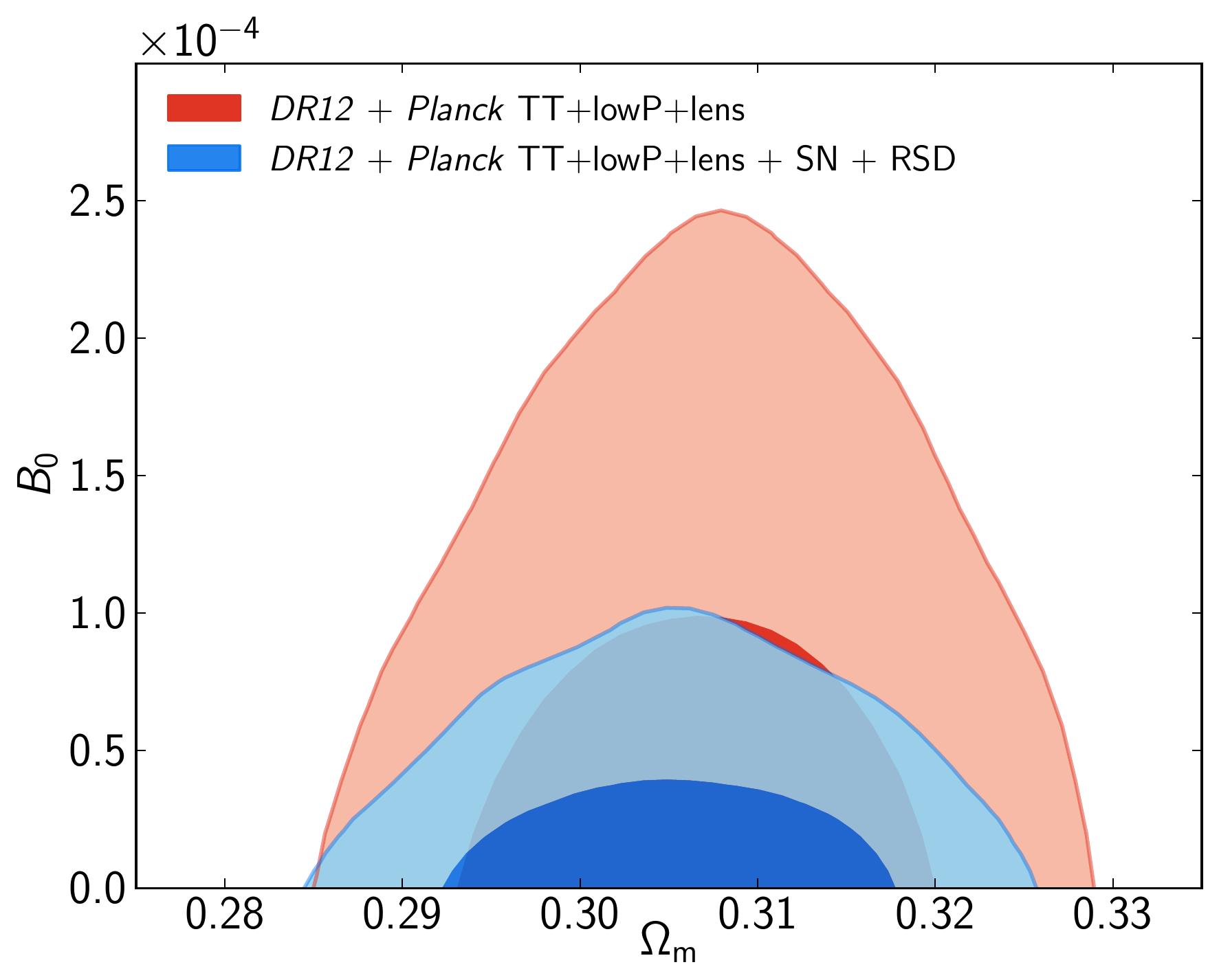}
  	\caption{ 2D contours of the $f(R)$ parameter $B_0$. The red contours refers to the DR12 combined analysis + Planck measurements ($B_0< 2.0 \times 10^{-4}$ at 95\% C.L.) whereas the blue contours include SN as well as additional RSD measurements ($B_0< 7.7 \times 10^{-5}$ at 95\% C.L.)}
	\label{fig:fR}
\end{figure}

%
Fig.~\ref{fig:fR} shows the constraints on $B_0$ in the one parameter $f(R)$ model as defined in Sec.~\ref{sec:BZ}. We find an upper limit of $B_0< 7.7 \times 10^{-5}$  at 95\% C.L including all data sets considered in this study. Neither model is favoured by the data compared to $\Lambda$CDM in our analysis.
Note, that we applied a linear prior on $B_0$ to sample its distribution function. Alternatively, one could assume a logarithmic prior on $B_0$ instead, to give equal weight to large and small scales. The caveat of this approach, however, is that the range of $\mrm{log}B_0$ is unknown a priori, introducing a dependence of the constraints on the lower limit of the prior. Since for all values of $\mrm{log}B_0<-6$, $f(R)$ mimics LCDM, we adopt a prior on $\mrm{log}B_0$ of [-6,-2] as in \citet{2015PhRvD..92d3522S}. We find an upper limit of $\mrm{log}B_0<-4.54$ at 95 \% C.L.. 

Another caveat in our analysis is that the BZ as well as $f(R)$ parametrisation is k-dependent whereas all RSD observations measure $f \sigma_8(z)$ at an effective scale of $k\approx0.15-0.2$ h/Mpc. Calculating $f \sigma_8(z)$ averaged over all scales as implemented in CosmoMC could potentially bias the results. The authors of \citep{Alam:2015rsa}, find that using the growth rate calculated at $k=0.2$ h/Mpc instead of averaged over k reduced the error on $B_0$ by 30-40\%. In general, a RSD measurement binned in redshift as well as scale $f(z,k)$ would be necessary to improve upon the errors on the BZ parameters and to detect a scale dependent deviation from gravity. We leave this analysis for future work.

\section{Conclusions}

In this paper, we have used recent galaxy clustering measurements made from the BOSS DR12 data to test for evidence supporting models that modify gravity beyond GR. We consider a number of extensions to the $\Lambda$CDM+GR model inspired by modifications to GR, and test whether these extensions are supported by the data. One of the simplest such model is the $\gamma$ parametrisation of the growth rate, which we introduced in Section~\ref{sec:growth_index}. In fact, we highlighted a subtlety in the common implementation of this model, in that people often combine $\gamma$ with additional assumptions, and we highlight two of these, which we call $\{ \gamma\ | \ \mrm{slip}\}$ and $\{ \gamma \ | \ G_{L} \}$, and we compare constraints from all three in Section~\ref{sec:Analysis_growth_index}. When comparing measurements made in different analyses, or in implementations in packages such as {\sc MGCAMB}, which uses the $\{\gamma \ | \ \mrm{slip}\}$ assumption, it is important to understand which model is being used. Moving beyond $\gamma$, we have introduced, in Section~\ref{sec:theory_GMGL}, dimensionless parameters $G_M$ and $G_L$ into the Einstein equations allowing non-GR evolution of the gravitational potentials $\phi$ and $\psi$. Finally, we consider a 5-parameter BZ parameterisation of deviations from GR as well as a one parameter $f(R)$ model, introduced in Section~\ref{sec:BZ}.

The BOSS DR12 measurements, along with those from the CMB are considered the most robust as they rely on simple physical processes and minimal additional modelling. The comparison between BOSS BAO and RSD measurements compares expansion and structure-growth, which is particularly powerful for making such measurements and testing GR, and the BOSS DR12 measurements are the most accurate to date. To extend the redshift range covered, we combine the BOSS measurements with BAO and RSD from the 6dFGS, VIPERS and the SDSS Main Galaxy Sample, chosen because they do not spatially overlap with BOSS. As well as the large-scale structure data, we include Planck CMB measurements, excluding polarisation data because of potential calibration issues, and the JLA supernovae data. These data sets were introduced in Section~\ref{sec:data_sets}.

Results from the fits to data are presented in Section~\ref{sec:Analysis}. For the $\gamma$ parameterisation, we see significant changes in the confidence intervals depending on the exact implementation: $\gamma$, $\{\gamma \ | \ \mrm{slip}\}$ or $\{ \gamma \ | \ G_{L} \}$, but all are consistent with GR. For the parameterisations with more free parameters we again see that the $\Lambda$CDM+GR model is an acceptable fit, showing no evidence requiring modified gravity.

Even though we have found no evidence requiring modifications to GR in the data sets analysed, there are a number of observations in mild tension with the simple $\Lambda$CDM+GR cosmological model. Given the free parameters within the $\Lambda$CDM+GR framework, these tensions generally show up when $\Lambda$CDM parameter measurements, made using different data, are compared. One source of tension is shown when the lensing measurements of the amplitude of matter clustering from CFHTLS \citep{Heymans:2012gg} and KiDS \citep{2016arXiv160605338H} are compared to those made with Planck CMB measurements including CMB-lensing, with the KiDS data showing a 2.3$\sigma$ tension with the Planck 2015 results \citep{2016arXiv160605338H}. There are also data sets where there is mild tension between measurements using the same probes: e.g. between the Planck 2015 results and those from combining WMAP, SPT, and ACT \citep{Hinshaw:2012aka, Story:2012wx, Sievers:2013ica, Calabrese:2013jyk}. In addition, several high-precision direct measurements of $H_0$ measure values about 10 per cent higher than those inferred from combinations of Planck and BOSS BAO data \citep{Riess:2011yx, Freedman:2012ny, Riess:2016jrr, 2016arXiv160703155A}. The DR12 analysis of \cite{2016arXiv160600439G} presents a 2.5$\sigma$ tension with the $\Lambda$CDM+GR model driven by measurements of the redshift-space bispectrum. While this level of tension is potentially interesting, it relies on modelling the redshift-space bispectrum, an less established field compared with modelling BAO and RSD measurements. In our analysis we have included measurements from 2-point clustering only, finding good consistency with the $\Lambda$CDM+GR model. None of these 'discrepancies' is at the level of providing strong evidence for a breakdown of the simple $\Lambda$CDM+GR model, and under-estimated systematic and/or statistical errors in one of more measurements cannot be ruled out at this stage. 

The recently reported tension of $f\sigma_8$ measurements from RSD measurements being lower than $\Lambda$CDM+GR expectations \citep{Macaulay:2013swa} has been alleviated by the recent BOSS DR12 results, which are within $1\sigma$ of the expectation \citep{2016arXiv160703155A}. Our work using these data and other to look for evidence of modified GR further supports the view that there is no remaining tension in the RSD measurements. 

\section*{Acknowledgements}
EM would like to thank Rachel Bean and Levon Pogosian for helpful discussions.
EM and WJP acknowledge support from the European Research Council through the Darksurvey grant 614030. WJP also acknowledges support from the UK Science and Technology Facilities Council grant ST/N000668/1 and the UK Space Agency grant ST/N00180X/1. EL thanks the Aspen Center for Physics, which is supported by NSF grant PHY-1066293, for a motivating environment. This work is supported in part by the Energetic Cosmos Laboratory and by the U.S. Department of Energy, Office of Science, Office of High Energy Physics, under Award DE-SC-0007867 and contract no. DE-AC02-05CH11231. AGS acknowledges support from the Trans-regional Collaborative Research Centre TR33 `The Dark Universe' of the German Research Foundation (DFG). FB acknowledges support from the UK Space Agency through grant ST/N00180X/1. GBZ is supported by the NSFC Grant No. 11673025, by the Strategic Priority Research Program ``The Emergence of Cosmological Structures" of the Chinese Academy of Sciences Grant No. XDB09000000, and by University of Portsmouth. SA is supported by the European Research Council through theCOSFORM Research Grant (\#670193).

Funding for SDSS-III has been provided by the Alfred P. Sloan Foundation, the Participating Institutions, the National Science Foundation, and the U.S. Department of Energy Office of Science. The SDSS-III web site is http://www.sdss3.org/.

SDSS-III is managed by the Astrophysical Research Consortium for the Participating Institutions of the SDSS-III Collaboration including the University of Arizona, the Brazilian Participation Group, Brookhaven National Laboratory, Carnegie Mellon University, University of Florida, the French Participation Group, the German Participation Group, Harvard University, the Instituto de Astrofisica de Canarias, the Michigan State/Notre Dame/JINA Participation Group, Johns Hopkins University, Lawrence Berkeley National Laboratory, Max Planck Institute for Astrophysics, Max Planck Institute for Extraterrestrial Physics, New Mexico State University, New York University, Ohio State University, Pennsylvania State University, University of Portsmouth, Princeton University, the Spanish Participation Group, University of Tokyo, University of Utah, Vanderbilt University, University of Virginia, University of Washington, and Yale University.


\bibliographystyle{mnras}
\bibliography{DR12_MG_ads}



\appendix


\bsp	
\label{lastpage}
\end{document}